\documentclass[journal]{IEEEtran}
\usepackage{lineno,hyperref}

\usepackage{amsmath}
\usepackage{amssymb}
\usepackage{booktabs}
\usepackage{graphicx}
\usepackage{bm}
\usepackage{multirow}
\usepackage{booktabs}
\usepackage[table,xcdraw]{xcolor}
\usepackage{pifont}

\usepackage{algorithm}
\usepackage{algpseudocode}

\begin{document}

	\title{FaultSSL: Seismic Fault Detection via Semi-supervised learning}
	\author{Yimin Dou, Minghui Dong, Kewen Li, Yuan Xiao
		\thanks{\textit{Corresponding author: Kewen Li}}
		\thanks{Yimin Dou, Minghui Dong, Kewen Li, Yuan Xiao College of computer science and technology, China University of Petroleum (East China) Qingdao, China.}
		\thanks{This work was supported by grntsa from the National Natural Science Foundation of China (Major Program, No.51991365), the Natural Science Foundation of Shandong Province, China (ZR2021MF082), and the Graduate Innovation Fund of China University of Petroleum (East China) (23CX04033A).}
	}
	\maketitle

\begin{abstract}

The prevailing methodology in data-driven fault detection leverages synthetic data for training neural networks. However, it grapples with challenges when it comes to generalization in surveys exhibiting complex structures.
To enhance the generalization of models trained on limited synthetic datasets to a broader range of real-world data, we introduce FaultSSL, a semi-supervised fault detection framework.
This method is based on the classical mean teacher structure, with its supervised part employs synthetic data and a few 2D labels. The unsupervised component relying on two meticulously devised proxy tasks, allowing it to incorporate vast unlabeled field data into the training process.
The two proxy tasks are PaNning Consistency (PNC) and PaTching Consistency (PTC). 
PNC emphasizes the feature consistency of the overlapping regions between two adjacent views in predicting the model. This allows for the extension of 2D slice labels to the global seismic volume.
PTC emphasizes the spatially consistent nature of faults. It ensures that the predictions for the seismic, whether made on the entire volume or on individual patches, exhibit coherence without any noticeable artifacts at the patch boundaries. 
While the two proxy tasks serve different objectives, they uniformly contribute to the enhancement of performance. 
Experiments showcase the exceptional performance of FaultSSL. In surveys where other mainstream methods fail to deliver, we present reliable, continuous, and clear detection results. FaultSSL breaks the shackles of synthetic data, unveiling a promising route for incorporating copious amounts of field data into training and fostering model generalization across a broader spectrum of surveys.
\end{abstract}

\section{Introduction}
Fault detection is a crucial step in oil and gas exploration. Traditional methods rely on anisotropy\cite{ruger1997p,ruger1997using}, coherence\cite{marfurt19983,bahorich19953,gersztenkorn1999eigenstructure}, ant tracking\cite{chopra2014seismic,sun2011application}, and other techniques, which are sensitive to hyperparameters and noise. This work focuses on Data-driven methods. 

\subsection{Challenge of Data Driven Fault Detection}
The difficulty of obtaining accurate labels has always been a serious problem in data-driven fault detection tasks. The complexity of the 3D distribution of faults makes it almost impossible to manually acquire seismic complete labels. 
The 2D labels are relatively easy to obtain, but 2D fault detection has inherent flaws, such as the discontinuity of 2D stitching results in three-dimensional space, the inability to obtain more sufficient spatial and neighborhood information, etc.\cite{zhao2018fault,guo2018new,di2020accelerating}. Therefore, we focus on the efforts made by researchers in 3D fault detection.

Guitton et al. build feature vectors for the training and classification steps using two popular techniques in object recognition algorithms called Histograms of Oriented Gradients and Scale Invariant Feature Transforms. The seismic volume is then classified point by point using the support vector machine \cite{guitton2017statistical}. A similar approach was taken by Di et al. who proposed a support vector machine method based on multiple attributes, which classifies using gradients, textures, amplitudes, etc.\cite{di2017seismic}. Xiong et al. treated seismic fault detection as an image classification task, taking the three sections near the center point of the seismic cube (inline, crossline, timeline) as input and outputting the fault probability of the center point \cite{xiong2018seismic}. Although these methods are ultimately applied to 3D seismic volumes, they are still 1D\cite{guitton2017statistical,di2017seismic} or 2D\cite{xiong2018seismic} methods.
Guitton was the first researcher to use 3D CNNs \cite{guitton20183d}. Although this method did not apply the image segmentation paradigm, its advantage over lower-dimensional methods is its ability to learn more spatial information. The above methods have two obvious drawbacks. First, the effectiveness of the model depends on the quality of the interpretation; second, point-by-point classification consumes a large amount of computational resources. However, given the difficulty of obtaining labels at the time, point-by-point classification was a feasible solution for the extremely small and sparse labels available.

To address the difficulty of obtaining 3D fault labels, Wu proposed a synthetic data method\cite{wu2019faultseg3d}. Subsequently, a large number of methods based on synthetic data emerged, most of which aimed at improving the backbone network and loss function \cite{wu2019faultseg3d,feng2021uncertainty,gao2022automatic,alohali2022automated,han2023algorithm,kaur2023deep,ma20233d,10196009,gao2021fault,li2022automatic}.
However, this also brings forth a potential issue, which is the inability to ensure that models trained on synthetic data can generalize to various complex types of surveys \cite{dou2021attention}. Currently, methods based on synthetic data have shown great promise in shallow and simple seismic volumes, but they lack reliability in deep and complex data.

Efforts are underway to mitigate the limitations of synthetic data. For example, researchers have investigated the use of manually labeled 2D slices for supervising 3D data \cite{dou2021attention}. However, manual labeling is error-prone and can adversely affect model performance. Therefore, some work  aims to minimize these impacts as much as possible \cite{dou2022md}.
These methods face two primary challenges. Firstly, in fault-rich regions, it is recommended to have a labeling proportion greater than 3.3\% \cite{dou2021attention}. This requirement imposes a significant workload, particularly in larger survey areas. Moreover, as the goal shifts towards training on more diverse and complex surveys, the labeling difficulty further escalates. Secondly, due to the inherent characteristics of faults, achieving completely accurate labeling is unattainable. With a vast amount of field data, the more labeling is conducted, the greater the accumulation of errors. Even with techniques aimed at minimizing the impact of incorrect labels, once errors reach a certain threshold, they can have a detrimental effect on model performance.

\subsection{Contribution}
To address the challenges posed by synthetic data and 2D slice-based approaches described above, we propose FaultSSL, a semi-supervised fault detection framework. It incorporates the strengths of both methods while overcoming the limitations faced by each of them. 
FaultSSL utilizes training samples from the synthetic seismic data provided by Wu's open-source dataset\cite{wu2019faultseg3d}, along with a substantial amount of field data. Each field data instance contains 1-3 annotated 2D slices. The information embedded in these slices is propagated throughout the entire seismic volume through the proxy task proposed by our method. Specifically, our contribution is as follows:

\subsubsection{PaNning Consistency (PNC)}
The PNC proxy task emphasizes that the overlapping regions between two adjacent views in seismic data should have identical predictions. Based on this premise, we can achieve two objectives:

Firstly, label propagation. We only annotate 1-3 slices for each field data, which encompass human-experienced feature descriptions of the survey. However, this is insufficient to represent the entire seismic volume. To address this, we employ the PNC proxy task, starting from the annotated slice locations and gradually propagating human experience throughout the whole region using overlapping areas as the medium for learning.

Secondly, feature consistency regularization.
Consistency regularization is a classic semi-supervised method that assumes the model should produce consistent features or labels for the same sample under different forms of data augmentation. PNC leverages consistency regularization by applying different augmentations to the two adjacent views. By doing so, it performs consistency regularization semi-supervised learning on field data using their overlapping regions as the medium. In practical implementation, it is only feasible to compute sparse overlapping regions. Therefore, PNC relies on the cosine loss to ensure feature consistency.

\subsubsection{PaTching Consistency (PTC)}
The assumption of the PTC proxy task is that, for the entire seismic volume, consistent predictions should be made regardless of whether it is done through cube-wise inference or holistic inference.
Building upon this premise, we can achieve two objectives:

Firstly, inference size-free.
In reality, cube-wise predictions often exhibit noticeable artifacts at the boundaries, and cube-wise inference is sometimes inevitable due to limitations in computer memory. These artifacts are caused by inconsistent receptive fields of convolutions for cubes of different sizes.
To address this, we randomly crop a cube from the field data and then randomly extract a subcube of random size within that cube. PTC enforces complete consistency between the predictions of the cube and subcube, this significantly reduces the generation of stitching artifacts.

Secondly, label consistency regularization. In PTC, the overlap between the cube and subcube is well-defined, allowing us to apply label consistency regularization to the output labels.

\subsubsection{FaultSSL}
FaultSSL is built upon the Mean Teacher architecture and thoughtfully designed training procedures, taking into account the specific characteristics of seismic data. It seamlessly integrates supervised learning and unsupervised PNC and PTC approaches. This integration breaks the limitations and closed nature of synthetic data, successfully addressing the challenge of accumulating errors in manual data annotations due to a large labeling workload. FaultSSL achieves state-of-the-art performance.

\subsubsection{Data augmentation}
We propose a set of data augmentation techniques specifically designed for fault detection tasks in seismic data. With the application of these data augmentation methods, our baseline model (trained on 200 synthetic data samples using HRNet) outperforms the majority of mainstream methods on the F3 and Kerry datasets.

\section{Approach}
\subsection{Problem Definition}
Synthetic data is represented as $\textbf{X}_\text{syn}=\{(\mathcal{X}_i,\mathcal{Y}_i):i\in(0,...\mathcal{K}) \}$, where $\mathcal{X}_i$ represents the $i^\text{th}$ cube, and $\mathcal{Y}_i$ is the corresponding 3D label.
Field data is represented as $\textbf{X}_\text{fld}=\{ (\mathcal{D}_i,y_{i,j}):i\in (0,...\mathcal{I}), j\in(0,...\mathcal{J}) \}$. Where $\mathcal{D}_i$ represents the $i^\text{th}$ 3D field data, and $y_{i,j}$ represents the 2D label of the  $j^\text{th}$ slice of the corresponding field data. It is important to note that  $y_{i,j}$ may not be entirely accurate.
This semi-supervised task aims to empower the network by leveraging a limited number of 2D slice labels $y_{i,j}$ from the field data (approximately 1-3 slices per survey) and combining them with synthetic data during training. This approach allows the network to achieve outstanding performance not only on the specific $\textbf{X}_\text{fld}$ dataset but also to effectively generalize its capabilities to a wider array of real-world data.

\subsection{Preliminary}
Figure \ref{framework} illustrates the complete framework of FaultSSL. To better convey our innovation, this section begins by introducing some published components employed in FaultSSL, as follows:
\subsubsection{Mean Teacher}
FaultSSL employs a dual-network framework, comprising of a student network (gradient network) and a teacher network (Exponential Moving Average (EMA) network) \cite{tarvainen2017mean}. This structure is widely used in current semi-supervised baseline frameworks. The student network is constructed based on gradient updates generated at each iteration, whereas the teacher network is updated according to the equation (\ref{ema}):
\begin{equation}
	\mathcal{V}_t^\text{ema} =\lambda \mathcal{V}_{t-1}^\text{ema}+(1-\lambda) \mathcal{V}_t
	\label{ema}
\end{equation}
Where $\mathcal{V}_t^\text{ema}$ represents the teacher network parameter at the current step, $\mathcal{V}_{t-1}^\text{ema}$ represents the teacher network parameter at the previous step, and $\mathcal{V}_t$ represents the  student network parameter at the current step. The momentum weight denoted by $\lambda$, satisfies the equation (\ref{lamda}).
\begin{equation}
	\lambda = 1-0.05\times (\textbf{\text{F}}_\text{cos}(\pi \mathcal{T}_\text{cu}/ \mathcal{T}_\text{all})+1)/2
	\label{lamda}
\end{equation}
$\mathcal{T}_\text{all}$ is the total number of steps to be trained, $\mathcal{T}_\text{cu}$ is the current step.

\subsubsection{HRNet}
In our framework, we have chosen HRNet\cite{wang2020deep} as the backbone due to its impressive performance in segmentation tasks. However, it's important to note that the specific structure of the network is not a critical aspect and can be replaced with any other segmentation network if desired.

\subsubsection{MD loss}
As mentioned in the introduction, we need to use region-based losses like Dice for our framework. This loss should also be applicable for training sparse labels. Dou et al. proposed the MD loss\cite{dou2022md}, which satisfies this requirement, defined as equation (\ref{md}).

\begin{equation}
	\mathcal{L}_{\text{md}}=1-\frac{2|\mathcal{M}(\hat{\mathcal{Y}})\cap \mathcal{M}(\mathcal{Y})|}{|\mathcal{M}(\hat{\mathcal{Y}})|+|\mathcal{M}(\mathcal{Y})|}\label{md}
\end{equation}
Where $\mathcal{Y}$ represents the ground truth,$\mathcal{\hat{Y}}$ denotes the predicted value, and $\mathcal{M}(\cdot)$ is the mask function used for unannotated regions.

\subsubsection{Cosine loss}
In PNC, we utilized the cosine loss, which is defined as shown in equation  (\ref{cosim}).
\begin{equation}
	\mathcal{L}_\text{cosim} = \frac{v_1\cdot v_2}{\text{max}(\parallel v_1 \parallel \cdot\parallel v_2 \parallel,\epsilon)} \label{cosim}
\end{equation}
where $\epsilon$ aim to avoid dividing by zero, $v_1$ and $v_2$ are the feature vectors.

The following content in this section presents our innovation and contributions.

\begin{figure*}[!h]
	\centering
	\includegraphics[scale=1.0]{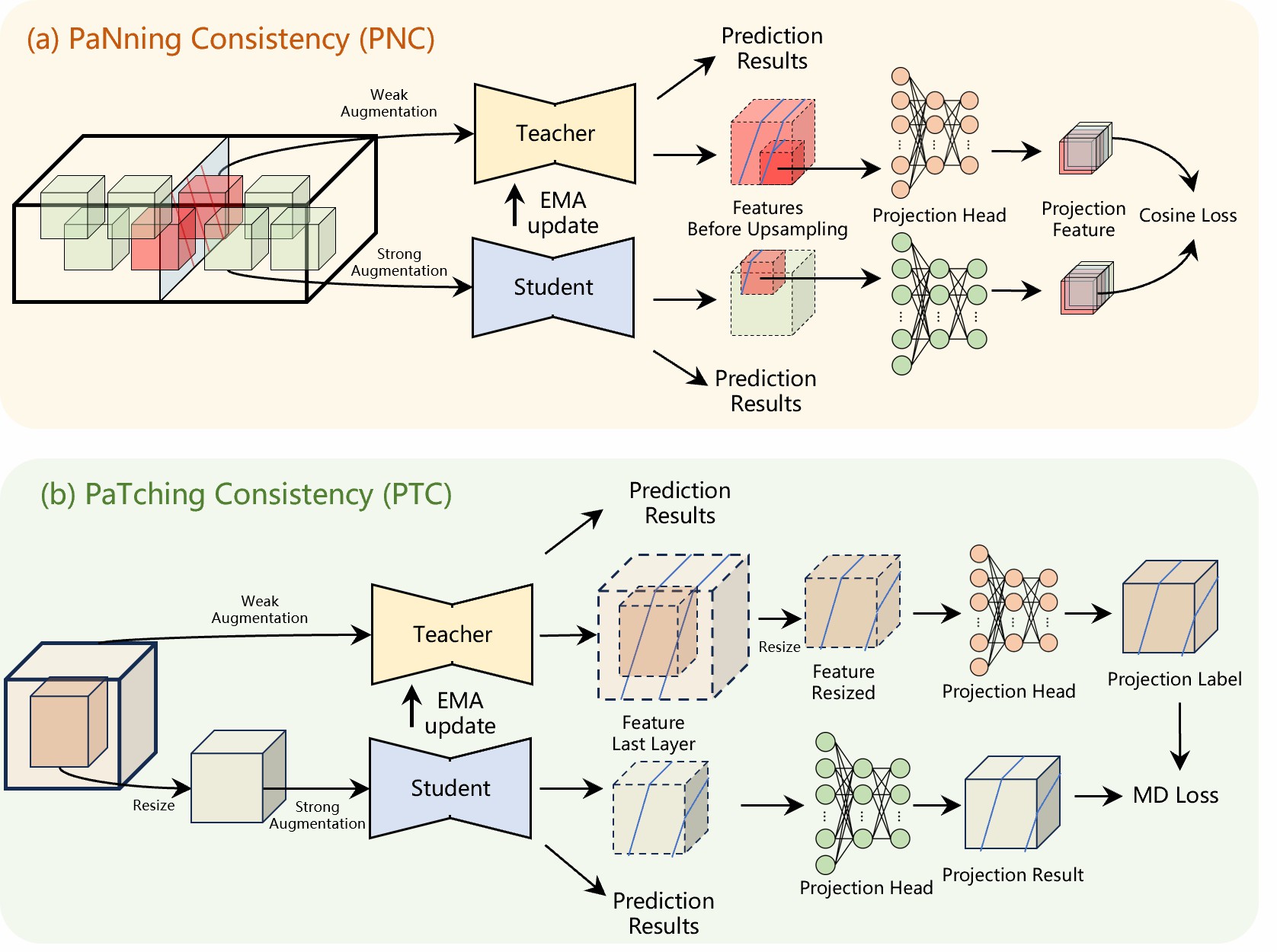}
	\centering\caption{FaultSSL's PNC and PTC proxy task flowchart.}
	\label{framework}
\end{figure*}

\subsection{PaNning Consistency (PNC)\label{pnc}}
Figure 1(a) illustrates the PNC proxy task.
The PNC proxy task assumes that the network has undergone extensive pretraining on synthetic data in a strong augmentation mode (refer to Part 3 of this section). As a result of this pretraining, the network can learn the fundamental patterns and morphologies of fault structures. This enables us to transfer the knowledge to field data with very few human prompts.

The field seismic data is represented as $\mathcal{D}_i \in \mathbb{R}^{l_\text{t} \times l_\text{i}\times l_\text{x}}$, and the labels for a slice in the inline direction are denoted as $y_{i,j} \in \mathbb{R}^{l_\text{t}\times l_\text{x}}$. The coordinates of this slice in the inline direction are represented as $p$. Let the side length of the training sample cube be denoted as $\l_\text{c}$. Then, each training sample of the field data can be represented as $\mathcal{X}_{\mathcal{D},i} \in \mathbb{R}^{\l_\text{c}\times\l_\text{c} \times\l_\text{c}}$, while the samples cropped from the seismic data at their original size are represented as $\hat{\mathcal{X}}_{\mathcal{D},i} \in \mathbb{R}^{\varphi_\text{t} l_\text{c} \times \varphi_\text{i} l_\text{c} \times \varphi_\text{x} l_\text{c}}$. 
The relationship between $\mathcal{X}_{\mathcal{D},i}$ and $\hat{\mathcal{X}}_{\mathcal{D},i}$ can be expressed by the equation (\ref{x2x}).

\begin{equation}
 \mathcal{X}_{\mathcal{D},i}= \textbf{\text{F}}^{\text{intp}}(\hat{\mathcal{X}}_{\mathcal{D},i}, (1/\varphi_\text{t},1/\varphi_\text{i},1/\varphi_\text{x}))
\label{x2x}
\end{equation}
Here, $\textbf{\text{F}}^{\text{intp}}(\cdot,\cdot)$ represents a trilinear interpolation function, where the first parameter is the input data, and the second parameter is the interpolation ratio, where,  $\varphi_\text{t},\varphi_\text{i},\varphi_\text{x} \sim \mathcal{U}(0.5,2)$.

We randomly crop samples from the field data, and a sample is determined by its center coordinates and corresponding side lengths $(\varphi_\text{t}l_\text{c},\varphi_\text{i}l_\text{c},\varphi_\text{x}l_\text{c})$ relative to the field data. The coordinates are represented by equation (\ref{cood}).

\begin{equation}
	\begin{aligned}
	&p_\text{t} \sim \mathcal{U}(\varphi_\text{t}l_\text{c}/2,l_\text{t}-\varphi_\text{t}l_\text{c}/2) \\
	&p_\text{x} \sim \mathcal{U}(\varphi_\text{x}l_\text{c}/2,l_\text{x}-\varphi_\text{x}l_\text{c}/2) \\
	p_\text{i} \sim \mathcal{U}(p(&1-\mathcal{T}_\text{cu}/\mathcal{T}_\text{all}),p-\varphi_\text{i}l_\text{c}+(l_\text{i}-p)(\mathcal{T}_\text{cu}/\mathcal{T}) )
	\end{aligned}
	\label{cood}
\end{equation}

The coordinates of $p_\text{t}, p_\text{i}, p_\text{x}$ represent the center  coordinates of $\hat{\mathcal{X}}_{\mathcal{D},i}$, and $p$ represents the position of the annotation slice along the inline, $l_\text{c}=128$.
To ensure the propagation of label information to a more distant extent, it is necessary to generate another sample, $\hat{\mathcal{X}}_{\mathcal{D},i+1}$, that satisfies the following two requirements. Firstly, the distance between the central coordinates of $\hat{\mathcal{X}}_{\mathcal{D},i+1}$ and the slice should be greater than the distance between $\hat{\mathcal{X}}_{\mathcal{D},i}$ and the slice. Secondly, $\hat{\mathcal{X}}_{\mathcal{D},i}$ and $\hat{\mathcal{X}}_{\mathcal{D},i+1}$ should guarantee the presence of at least an $l_\text{c}^3/8$ overlapping area.  

\begin{equation}
	\begin{aligned}
		&p_{\text{t},1} \sim \mathcal{U}(p_\text{t}-\varphi_\text{t}l_\text{c}/2,p_\text{t}+ \varphi_\text{t}l_\text{c}/2) \\
	&p_{\text{x},1} \sim \mathcal{U}(p_\text{x}-\varphi_\text{x}l_\text{c}/2,p_\text{x}+ \varphi_\text{x}l_\text{c}/2) \\
		&p_{\text{i},1} \sim 
		\begin{cases}
		\mathcal{U}( p_\text{i},p_\text{i}+\varphi_\text{t}l_\text{c}/2) \ \  \text{if}\  p_i-p>0 \\
		\mathcal{U}(p_\text{i}-\varphi_\text{t}l_\text{c}/2, p_\text{i}) \ \  \text{else}
		\end{cases} 
	\end{aligned}
	\label{cood1}
\end{equation}
To obtain $\mathcal{X}_{\mathcal{D},i}$ and $\mathcal{X}_{\mathcal{D},i+1}$, both $\hat{\mathcal{X}}_{\mathcal{D},i}$ and $\hat{\mathcal{X}}_{\mathcal{D},i+1}$ should be interpolated to the size of $l_\text{c}\times l_\text{c}\times l_\text{c}$ by equation (\ref{x2x}).

$\mathcal{X}_{\mathcal{D},i}$ is subjected to weak augmentation, while  $\mathcal{X}_{\mathcal{D},i+1}$ is subjected to strong augmentation.
Input $\mathcal{X}_{\mathcal{D},i}$ into the teacher network and $\mathcal{X}_{\mathcal{D},i+1}$ into the student network. 
Based on Dou et al.'s method\cite{dou2023contrasinver}, calculate the coordinates of the overlapping region after downsampling by a factor of 4 using HRNet. Then, extract the corresponding feature vectors of the overlapping region before upsampling. Input these vectors into their respective projection heads and calculate the cosine loss (equation (\ref{cosim})) of their outputs.

By repeating the above steps, as the value of $\mathcal{T}_\text{cu}$ increases in equation (\ref{cood}), the  2D slice label's information progressively propagates to a broader global context.

\subsection{PaTching Consistency (PTC)\label{ptc}}

Generate the original PTC sample $\hat{\mathcal{X}}_{\mathcal{D},j}$ based on equation (\ref{cood}), with the parameter $l_\text{c}=192$. Then, extract a subcube from $\hat{\mathcal{X}}_{\mathcal{D},j+1}$ according to equation (\ref{cood2}).
\begin{equation}
	\begin{aligned}
		&l_\text{c,1}\sim \mathcal{U}(\varphi_\text{t}l_\text{c}/2, \varphi_\text{t}l_\text{c})\\
		&l_\text{c,2}\sim \mathcal{U}(\varphi_\text{i}l_\text{c}/2, \varphi_\text{i}l_\text{c})\\
		&l_\text{c,3}\sim \mathcal{U}(\varphi_\text{x}l_\text{c}/2, \varphi_\text{x}l_\text{c})\\
		&\rho_\text{t} \sim \mathcal{U}(l_\text{c,1}/2, \varphi_\text{t}l_\text{c}-l_\text{c,1}/2) \\
		&\rho_\text{i} \sim \mathcal{U}(l_\text{c,2}/2, \varphi_\text{i}l_\text{c}-l_\text{c,2}/2) \\
		&\rho_\text{x} \sim \mathcal{U}(l_\text{c,3}/2, \varphi_\text{x}l_\text{c}-l_\text{c,3}/2)
	\end{aligned}
	\label{cood2}
\end{equation}
Where $l_\text{c,1}$, $l_\text{c,2}$, and $l_\text{c,3}$ are the side lengths of the subcube, and $\rho_\text{t}$, $\rho_\text{i}$, and $\rho_\text{x}$ are the corresponding central coordinates.

Basing equation \ref{x2x}, interpolate  $\hat{\mathcal{X}}_{\mathcal{D},j}$ and  $\hat{\mathcal{X}}_{\mathcal{D},j+1}$ to sizes $192\times192\times192$ and $128\times128\times128$, respectively, and denote them as $\mathcal{X}_{\mathcal{D},j}$  and $\mathcal{X}_{\mathcal{D},j+1}$.

As depicted in Figure \ref{framework}, $\mathcal{X}_{\mathcal{D},j}$ undergoes weak augmentation and is input to the teacher network, while $\mathcal{X}_{\mathcal{D},j+1}$ undergoes strong augmentation and is fed into the student network. The final layer features output by the teacher network undergo the same operation as equation \ref{cood2}. Then, these features are resized to the same size as the student network's output using nearest-neighbor interpolation before being passed through the projection head to obtain pseudo-labels. The output features of the student network are combined with the pseudo-labels to compute the MD loss.

PTC requires the network to make predictions that are entirely consistent for cubes with different scales or receptive fields. This constraint compels the network to learn edge features of the underlying structures in the seismic, resulting in smoother and more accurate predictions at the seismic cube's edges. By enforcing this consistency across various scales or receptive fields, PTC enhances the network's ability to capture lower-level features and improves its overall performance.

\subsection{Data Augmentation}
According to the requirements of the PNC and PTC proxy tasks, it is necessary to establish strong and weak augmentation methods. Although there are some augmentation techniques available in the domain of natural images, these methods are challenging to apply directly to seismic data. Therefore, we propose the following augmentation approaches specifically designed for seismic data.

\subsubsection{Weak Augmentation}
Weak augmentation includes only coordinate transformations.

\textit{Random flipping}: The input data is randomly flipped in three directions (timeline, inline, crossline).

\textit{0-600° rotation}: Existing methods mostly utilize rotations in right angles, thereby neglecting the significance of capturing additional directional structures in training seismic data. Specifically, we incorporate random rotations in the range of 0-45° for inline and crossline profile, as well as random rotations in the range of 0-30° for inline and timeline profile, in addition to the existing right-angle rotations and flips. These random non-right-angle rotations, combined with right-angle rotations and flips, constitute a data augmentation technique involving random rotations in the range of 0-600°.

\subsubsection{Strong Augmentation}
Strong augmentation includes not only the coordinate transformations mentioned above but also voxel-level and patch-level transformations.

\textit{Random-scale Gaussian noise}:
Incorporate random-scale Gaussian noise during training, as defined in equation (\ref{nsa}).
\begin{equation}
	\text{Aug}_\text{ns}(\mathcal{X}_i) = \mathcal{X}_i + \mathcal{G}, \ \mathcal{G} \thicksim \mathcal{N}(0,\sigma_\text{n}), \ \sigma_\text{n} \thicksim   \mathcal{U}(0.0,0.15) \label{nsa}
\end{equation}
Among them, $\mathcal{X}$ represents the input data, and $\mathcal{G}$ is random-scale Gaussian noise with variance following a uniform distribution. We recommend the range for the variance to be $\sigma \thicksim   \mathcal{U}(0.0,0.15)$. Regardless of the improvements made to the method, Gaussian noise augmentation is \textbf{indispensable}.

\textit{Random gamma transformation}: The expression for applying random gamma non-linear transformation to each voxel is given by the equation (\ref{gma}).
\begin{equation}
	\text{Aug}_\text{gm}(\mathcal{X}_i) = \mathcal{X}_i^\gamma , \ \gamma \thicksim \mathcal{N}(1,0.1) \label{gma}
\end{equation}

\textit{Random filtering}: Randomly adding Gaussian filtering at different scales is recommended, with $\sigma_{\text{f},d}  \thicksim  \mathcal{U}(0.1,0.6)$ ($\sigma_{\text{f},d}$ is not necessarily equal in all three directions). This is crucial for the network's ability to detect fine and small-scale features, such as subtle or tiny discontinuities in the data.

\subsection{Workflow}
The workflow of FaultSSL is summarised here as shown in Algorithm \ref{wfl}.
FaultSSL's pretraining involves a supervised process using complete labels of synthetic data. The semi-supervised training consists of three tasks: first, a supervised task where within a batch, half of the data is synthetic while the other half is field data with slice labels. Second, the PNC proxy task. Third, the PTC proxy task. In a single step, these three tasks are sequentially executed, and their accumulated gradients are used to update the student network. Then, the teacher network is updated via EMA.

\begin{algorithm}[!h]
		\scriptsize
		\caption{FaultSSL workflow}
		\label{wfl}
		\begin{algorithmic}[1]
			\State \textbf{Initialize:} Student model parameters $\theta$, Teacher model parameters $\Theta = \theta$
			\Procedure{Pretrain}{$\Theta ,\theta $, $\textbf{X}_\text{syn}$} \Comment{$\textbf{X}_\text{syn}=\{(\mathcal{X}_i,\mathcal{Y}_i):i\in(0,...\mathcal{K}) \}$}
			\For {each pretraining step}
			\State \textbf{Input} Sample a mini-batch from $\textbf{X}_\text{syn}$, get $\mathcal{X}_i,...\mathcal{X}_{i+b},\mathcal{Y}_i,...\mathcal{Y}_{i+b}$ \Comment{$b$ is batch size}
			\State Compute the prediction $\mathcal{\hat{Y}}_i,...\mathcal{\hat{Y}}_{i+b} = f_{\theta}(\mathcal{X}_i,...\mathcal{X}_{i+b})$ using Student model
			\State Compute $\mathcal{L}_\text{md}(\mathcal{Y}_i,...\mathcal{Y}_{i+b},\mathcal{\hat{Y}}_i,...\mathcal{\hat{Y}}_{i+b})$
			\State Student model backpropagation gradient
			\State Student updata $\theta$ by gradient
			\State Teacher updata $\Theta$ by EMA method
			\EndFor
			\EndProcedure
			\Procedure{Semi-supervised}{$\Theta, \theta $, $\textbf{X}_\text{syn}$,$\textbf{X}_\text{fld}$}
			\Comment{$\textbf{X}_\text{fld}=\{ (\mathcal{D}_i,y_{i,j}):i\in (0,...\mathcal{I}), j\in(0,...\mathcal{J}) \}$}
			\For {each semi-supervised step}
			\State \textbf{Input} Sample a mini-batch from $\textbf{X}_\text{syn}$, get $\mathcal{X}_i,...\mathcal{X}_{i+b/2},\mathcal{Y}_i,...\mathcal{Y}_{i+b/2}$. Sample a mini-batch from $\textbf{X}_\text{fld}$, get $\mathcal{X}_{\mathcal{D},k},...\mathcal{X}_{k+b/2}$ and $ y_k,...y_{k+b/2}$ 
			\State Get $\mathcal{\hat{Y}}_i,...\mathcal{\hat{Y}}_{i+b/2},\mathcal{\hat{Y}}_{\mathcal{D},k},...\mathcal{\hat{Y}}_{k+b/2} = f_{\theta}(\mathcal{X}_i,...\mathcal{X}_{i+b/2},\mathcal{X}_{\mathcal{D},k},...\mathcal{X}_{k+b/2})$ using student model
			\State Compute $\mathcal{L}_\text{md}(\mathcal{Y}_i,...\mathcal{Y}_{i+b/2},y_{\mathcal{D},k},...y_{k+b/2},
			\mathcal{\hat{Y}}_i,...\mathcal{\hat{Y}}_{i+b/2},\mathcal{\hat{Y}}_{\mathcal{D},k},...\mathcal{\hat{Y}}_{k+b/2})$
			\State Student model backpropagation gradient
			\State Execute PNC and calculate cosine loss \Comment{See section 2.2}
			\State Student model backpropagation gradient
			\State Execute PTC and calculate MD loss \Comment{See section 2.3}
			\State Student model backpropagation gradient
			\State Student updata $\theta$ by gradient \Comment{Gradient vectors from supervised, PNC,PTC, three tasks.}
			\State Teacher updata $\Theta$ by EMA method
			\EndFor
			\EndProcedure
		\end{algorithmic}
\end{algorithm}

\section{Experiments}
In the fault detection task, the quantitative results on synthetic data do not have a direct relationship with the qualitative performance on real data. Therefore, this study focuses only on qualitative experiments.

\subsection{Experimental Settings}
\subsubsection{Training data}
\textit{Synthetic data:} 
The synthetic data used 200 cubes from Wu\cite{wu2019faultseg3d} open source.

\textit{Field data:}
The field data used along with their corresponding annotations are presented in Table \ref{tab:t1}.
\begin{table}[!h]
	\setlength\tabcolsep{1pt}
	\centering
	\caption{The field data annotation status}
	\label{tab:t1}
	\begin{tabular}{@{}ccccccccc@{}}
		\toprule
		Survey      & Kerry &F3   & Poseidon & Canning & Opunake & CostaRica & Clyde & Niuzhuang \\ \midrule
		Labeled Num & 2     &2    & 2        & 3       & 2       & 1         & 1     & 2         \\ \bottomrule
	\end{tabular}
\end{table}

Other field data is also involved in the training, but they have not been labeled. Their PNC starting positions are set to the center of the survey. These unlabeled survey are Parihaka, Ichthys, Adele, and some data provided by Sinopec.

\subsubsection{Comparison Methods}

We added several control groups (ablation groups) to verify the effectiveness of each component of FaultSSL.
1) Baseline Model.
2) Baseline + Field Data Labeled Slices.
3) Baseline + Field Data Labeled Slices + PTC.
4) Baseline + Field Data Labeled Slices + PNC.
5) FaultSSL (Baseline + Field Data Labeled Slices + PTC + PNC).
Additionally, when displaying the results, we also included the results from Wu et al.'s FaultSeg3D\cite{wu2019faultnet3d} because all deep learning methods published after 2019 have referred to this work.

Additionally, the open-source code available for data-driven fault detection is very limited, which poses challenges for the reproduction of these works, and therefore makes it difficult to implement one-to-one qualitative comparisons. To overcome this difficulty, we have added the results of F3 and Kerry from all the papers published after 2019 in Geophysics and IEEE Transactions on Geoscience and Remote Sensing (TGRS) in the appendix (through screenshot), and we recommend reading the experiments in conjunction with the appendix(Figure \ref{apd1-1},\ref{apd1-2},\ref{apd2}). These methods include the following works: \cite{wu2019faultseg3d,feng2021uncertainty,dou2021attention,dou2022md,gao2022automatic,alohali2022automated,han2023algorithm,kaur2023deep,ma20233d,10196009,gao2021fault,li2022automatic}.

\subsubsection{Implementation Details}
FaultSSL is trained on two 3090ti GPUs using PyTorch 2.0 implementation. The optimizer used is AdamW with a learning rate of 0.001. The pretraining process is executed for 30,000 steps, while the semi-supervised training is performed for 160,000 steps.

For the baseline model, we employed HRNet with the same architecture, following the current mainstream approach of training based on synthetic data. Specifically, we utilized the 200 synthetic data samples provided by Wu et al. in their open-source dataset. Additionally, we applied the strong data augmentation techniques proposed in this paper for training process.

Due to memory constraints, our prediction results are generated by segmenting the seismic volume into blocks for prediction and then stitching them together. For some methods, this process can introduce stitching artifacts. Hereafter, this operation will be referred to as "cubing prediction."

\subsection{Comparison Experiment}

\subsubsection{F3 comparing experimental results.}
\begin{figure*}[!h]
	\centering
	\includegraphics[scale=0.275]{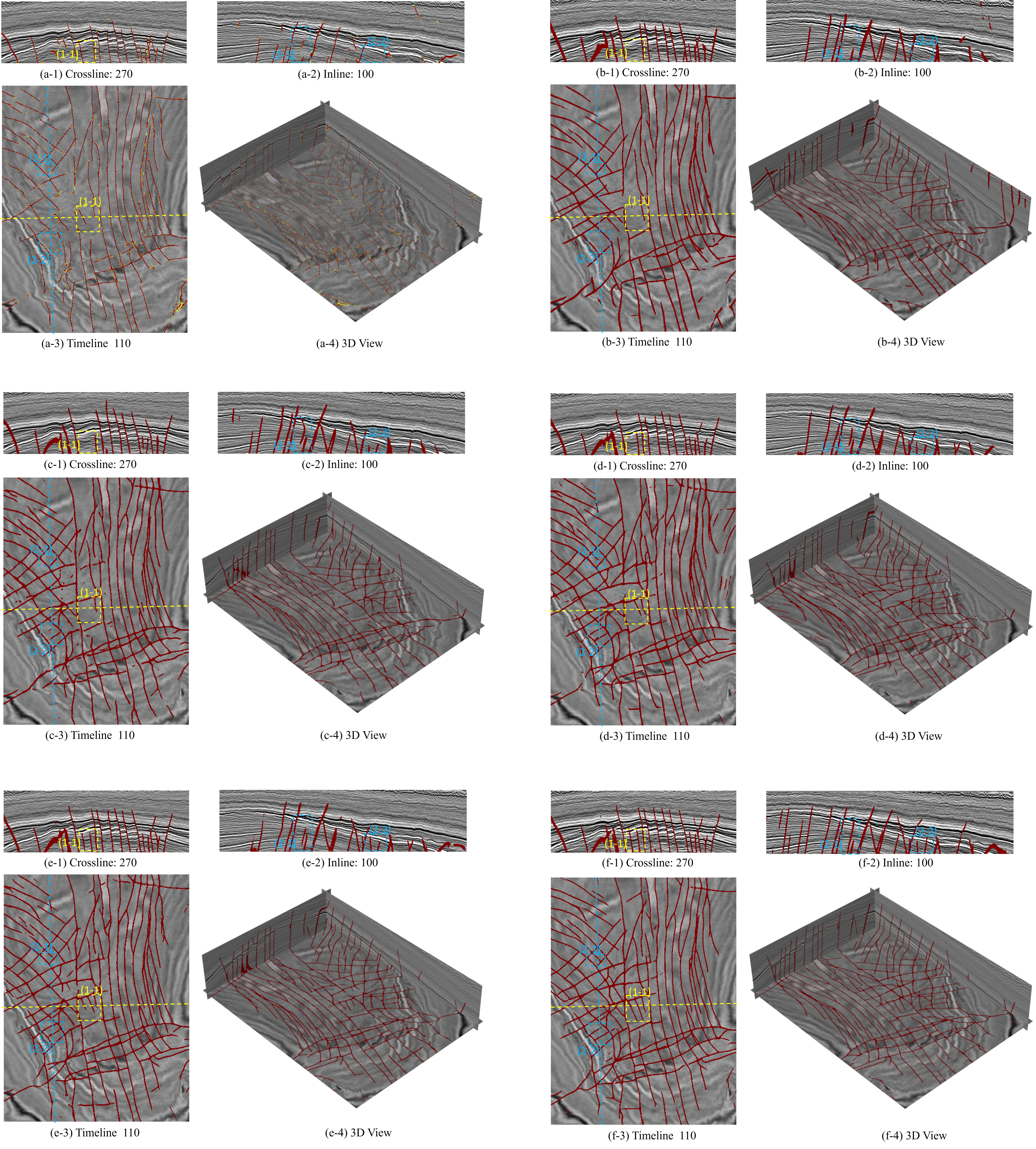}
	\centering\caption{(a) FaultSeg3D \cite{wu2019faultseg3d}. (b) Baseline Model. (c) Baseline + Field Data Labeled Slices. (d)  Baseline + Field Data Labeled Slices + PTC. (e)  Baseline + Field Data Labeled Slices + PNC. (f)  FaultSSL (Baseline + Field Data Labeled Slices + PTC + PNC).  
	Furthermore, in the appendix, we displayed the detection results for the F3 from all the papers published in Geophysics and TGRS after 2019 (Figure \ref{apd1-1}, \ref{apd1-2}).}
	\label{exp1}
\end{figure*}

Figure \ref{exp1} is a comparison of five sets of experiments and Wu et al.'s FaultSeg3D. FaultSeg3D was proposed by Wu et al. in 2019, and all deep learning methods have referenced this work since then. Unfortunately, although these methods all have their own highlights, very few works have significantly outperformed FaultSeg3D in terms of performance (see Figure \ref{apd1-1}, \ref{apd1-2} in the appendix). This suggests that the task of fault detection may not have made significant progress in the past four years.

First, we focus on the significant progress made in Figure \ref{exp1} (f) compared to (a), especially in the Timeline section. FaultSSL detected more faults, many of which have never been seen in all the studies of the F3 work area over the past four years. This forces us to question whether these faults really exist. Therefore, we selected three ROIs (regions of interest) in the Timeline, all of which were detected by FaultSSL and not by FaultSeg3D. We want to verify whether the faults in these three ROIs really exist. Since it is very difficult to observe faults from the Timeline section, we take the vertical sections (a-1), (a-2), (f-1), (f-2) corresponding to these three ROIs, and we mark the positions corresponding to the ROIs with boxes of the same color. From these vertical sections, we can easily observe that there are indeed faults at these locations.
Is the excellent performance of FaultSSL entirely due to manually annotated information? The answer is no. Although manual annotation is important, fully realizing its potential requires the assistance of the FaultSSL semi-supervised framework.
Figure \ref{exp1} (b), (c), (d), (e), and (f) demonstrate the effectiveness of each component of FaultSSL.

Figure \ref{exp1} (b) is the result of training entirely with synthetic data. Although this result has a gap compared to (f), it shows advantages over existing methods by applying the data augmentation method we proposed \cite{han2023algorithm}, \cite{kaur2023deep}, \cite{10196009}, \cite{feng2021uncertainty},\cite{dou2021attention} \cite{ma20233d}, \cite{alohali2022automated},  \cite{gao2022automatic}. (b) only applied the basic HRNet, and except for unique data augmentation, did not make any network improvements specifically for faults. We need to reflect on the significance of most of the improvement works after FaultSeg3D for the detection performance itself. 

Panel (c) includes 2D manual annotations, but does not use the assistance of the semi-supervised framework, so it can be considered a weakly supervised method. Although this method also showed progress, the detection of faults has obvious discontinuities, some faults were not detected completely, and compared to other methods using proxy tasks, it showed a lower recall. This is because the network only focuses on the two annotated 2D slices, and faults in other positions are ignored. This results in learning some ambiguous features. Although some obvious positions can be detected as faults, for some difficult areas, the network can only detect discontinuous faults.

Panel (d) uses the PTC proxy task. The essence of the PTC proxy task is scale consistent regularization, which can reduce the artifacts produced during cubing prediction, but the fault-rich area of F3 is small ($128\times 384 \times 512$), and it supports whole-volume inference, so here we only show and explain the performance improvement of PTC (the performance of handling artifacts in the Kerry survey is displayed). PTC implements a kind of consistency regularization semi-supervised method, therefore, its performance improvement is significant, and the results of (d) are more continuous compared to (c).

Panel (e) shows better results compared to (d) because of its diffusion supervision of sparse labels, maximizing the utilization of scarce labels as much as possible, and allowing supervised information to diffuse from local to global. The results become more continuous, and the performance improvement is significant.

Panel (f) is the result of FaultSSL, which integrates two kinds of proxy tasks, and its performance is significantly better than all current methods (see Appendix Figure \ref{apd1-1}, \ref{apd1-2}.).

\subsubsection{Kerry comparing experimental results.}

Scholars generally lean towards conducting fault detection in the shallow layers of the Kerry dataset, as its patterns resemble synthetic data, leading to more reliable outcomes.

\begin{figure*}[!h]
	\centering
	\includegraphics[scale=0.26]{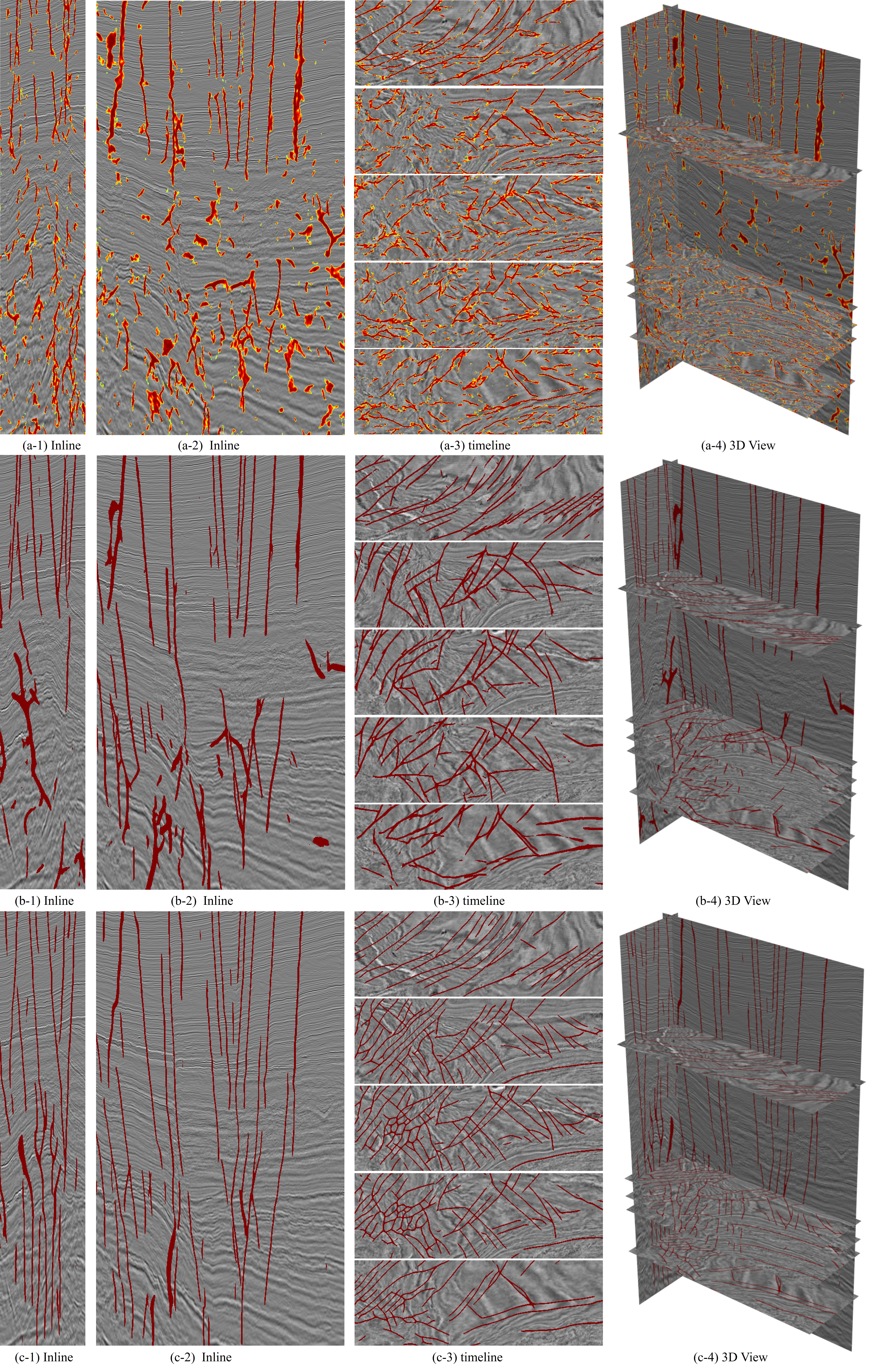}
	\centering\caption{(a) FaultSeg3D \cite{wu2019faultseg3d}.  (b) Baseline Model.  (c) FaultSSL (Baseline + Field Data Labeled Slices + PTC + PNC). Furthermore, in the appendix, we displayed the detection results for the Kerry from all the papers published in Geophysics and TGRS after 2019 (Figure \ref{apd2}).}
	\label{exp2}
\end{figure*}

Figure \ref{exp2} shows the results of FaultSeg3D, Baseline, and FaultSSL, respectively. Figure \ref{exp3} is a comparison of several typical sections. In Figure \ref{apd2} of the appendix, we have captured all the fault detection results for Kerry published after 2019 in Geophysics and TGRS. 

In Figure \ref{exp2} (a), we used the weights open-sourced by Wu in 2019 for the results of FaultSeg3D. Although a promising data-driven fault detection method (synthetic data) was explored at that time, the performance was far from satisfactory. In 2020, Wu's improved version (see Figure \ref{apd2} (a), not open-sourced) significantly improved the detection effects on Kerry. Most of the methods that emerged subsequently did not show a significant difference compared to Figure \ref{apd2} (a). FaultSeg3D and its derivative methods all share a common flaw, that is, they find it difficult to achieve stable generalization in the deep parts of seismic activities. Taking the deep part of Kerry as an example, the results of FaultSeg3D in Figure \ref{apd2}(a) show clearly disordered faults, and although (b) is cleaner and clearer than the results in (a), it misses a large number of faults, resulting in a lower recall. Since the deep layers are difficult to be forward modeled, and the fault morphology in the deep part is also quite special, there is currently no synthetic data available for the deep part. However, our method provides a completely new approach.
Due to the current limitations of synthetic data in providing fault information that encompasses the deeper regions of the Kerry dataset, as indicated in Table \ref{tab:t1}, we have annotated two inline slices of Kerry in order to facilitate robust generalization in its deeper layers. 
Figure \ref{exp2}(c) shows our results, which are far ahead of existing methods, especially in the depiction of deep faults. FaultSSL displays a clear, continuous, and precise Kerry fault detection result.

In Figure 4, we display several key sections to verify the effectiveness of each component of FaultSSL. 
(a) displays the results of FaultSeg3D.
(b) showcases the outcomes from synthetic data, which miss the majority of the faults. 
(c) represents results obtained by combining human-labeled slices without utilizing any proxy tasks. While it detects more faults, the results exhibit noticeable discontinuity.
(d) employs the PTC proxy task, detecting a higher number of faults, many of which are unlabeled. However, the precision remains suboptimal. Notably, at the location indicated by the yellow arrow, the continuity of the cubing predictions is optimal when compared to methods without PTC.
(e) employs PNC, which closely resembles FaultSSL but still exhibits cases of discontinuous cubing predictions.
(f) depicts the results from FaultSSL, showcasing the best performance.

Based on this study, the following conclusions can be drawn:
1) Incorporating human expertise yields superior fault detection results compared to using pure synthetic data.
2) The PTC proxy task provides a marginal performance improvement and mitigates the artifacts introduced by cubing predictions.
3) PNC significantly enhances performance, improving the continuity of prediction outcomes and maximizing the advantages of human annotations.
4) While PNC and PTC proxy tasks serve different purposes, both contribute to the enhancement of fault detection performance due to their effective utilization of unlabeled data.

\begin{figure*}[!h]
	\centering
	\includegraphics[scale=0.45]{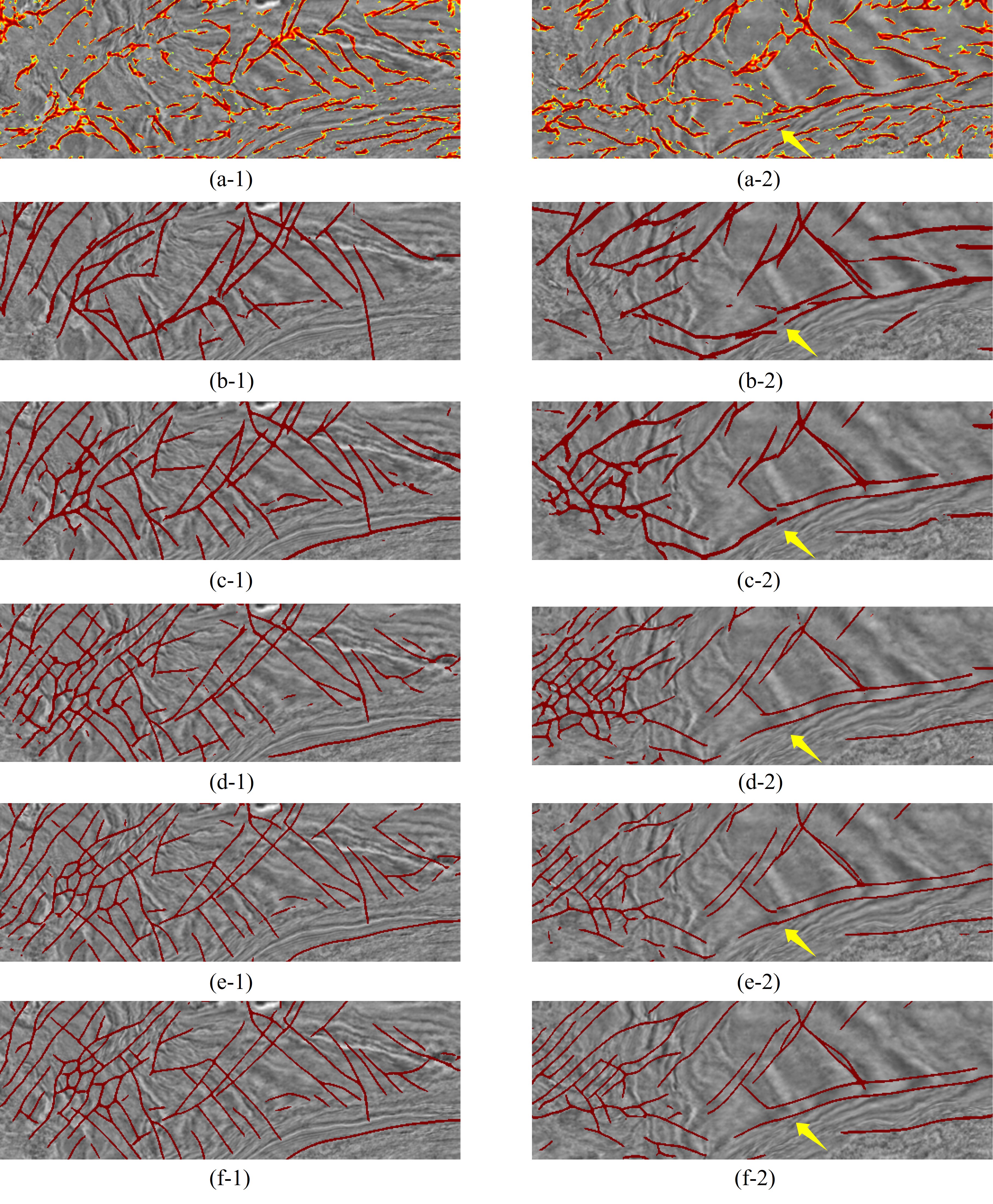}
	\centering\caption{(a)FaultSeg3D \cite{wu2019faultseg3d}. (b) Baseline Model. (c) Baseline + Field Data Labeled Slices. (d) Baseline + Field Data Labeled Slices + PTC. (e) Baseline + Field Data Labeled Slices + PNC. (f) FaultSSL.}
	\label{exp3}
\end{figure*}

\section{Conclusion}
In this paper, we present a semi-supervised fault detection method that utilizes two proxy tasks for combined training on synthetic data, manually annotated 2D slices, and a large amount of unlabeled data. Compared to the current methods based on synthetic data, the integration of onsite data with 2D labels can significantly enhance performance in specific surveys. Additionally, unsupervised training can improve the continuity of the detection results. Compared to other methods that introduce field data, FaultSSL greatly reduces the annotation workload, maximizing the use of a few slices, thus avoiding the performance decline caused by the excessive accumulation of manual slicing errors. Our experiments achieved the best qualitative results in two challenging work areas since 2019, and subsequent ablation studies also demonstrated the effectiveness of each component. FaultSSL outlines a method to rationally incorporate field data into training, achieving impressive results in surveys where synthetic data fails to generalize. This approach has the potential to break the current limitations of synthetic data-based techniques, allowing data-driven fault detection methods to adapt to more real-world data, bringing hope to the development of a universal fault detection model.

\section{Appendix}
\begin{figure*}[!h]
	\centering
	\includegraphics[scale=0.2]{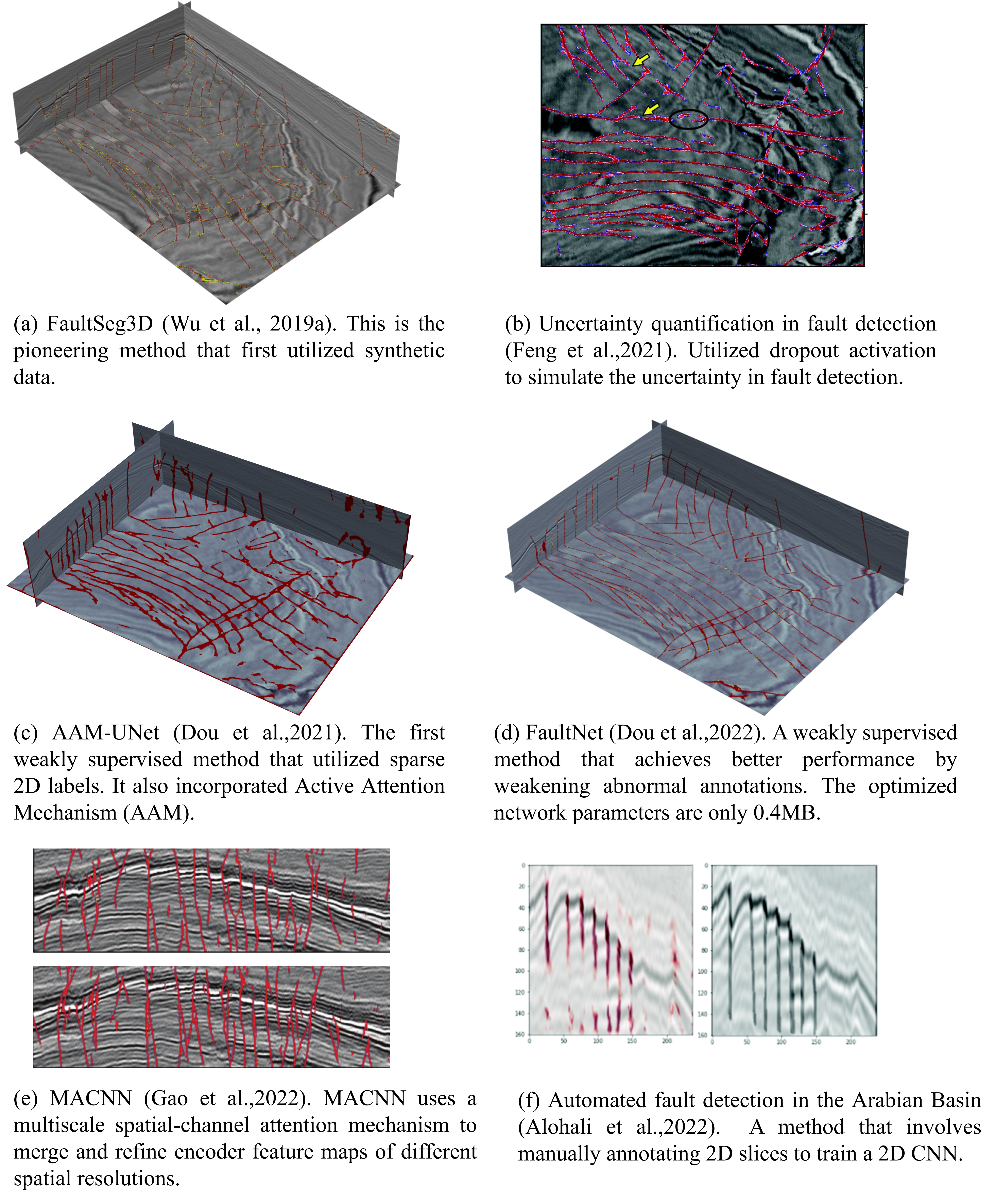}
	\centering\caption{Appendix figure: The detection results of the fault-rich area in the Netherlands F3 block from all the papers published in the two authoritative journals, Geophysics and TGRS, from 2019-2022. We summarized the main innovations of each method in one sentence.}
	\label{apd1-1}
\end{figure*}

\begin{figure*}[!h]
	\centering
	\includegraphics[scale=0.2]{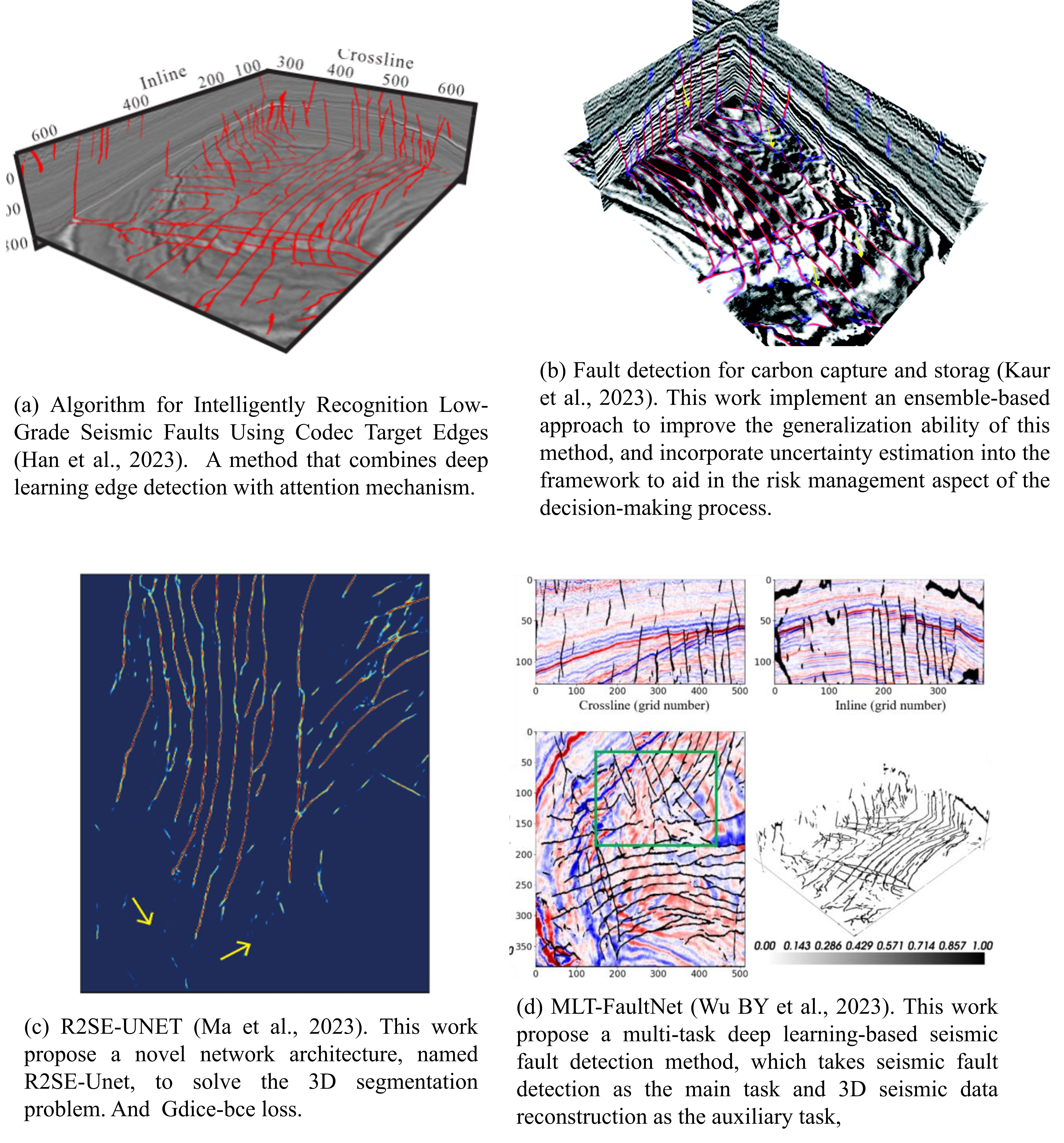}
	\centering\caption{Appendix figure: The detection results of the fault-rich area in the Netherlands F3 block from all the papers published in the two authoritative journals, Geophysics and TGRS, in 2023. We summarized the main innovations of each method in one sentence.}
	\label{apd1-2}
\end{figure*}

\begin{figure*}[!h]
	\centering
	\includegraphics[scale=0.165]{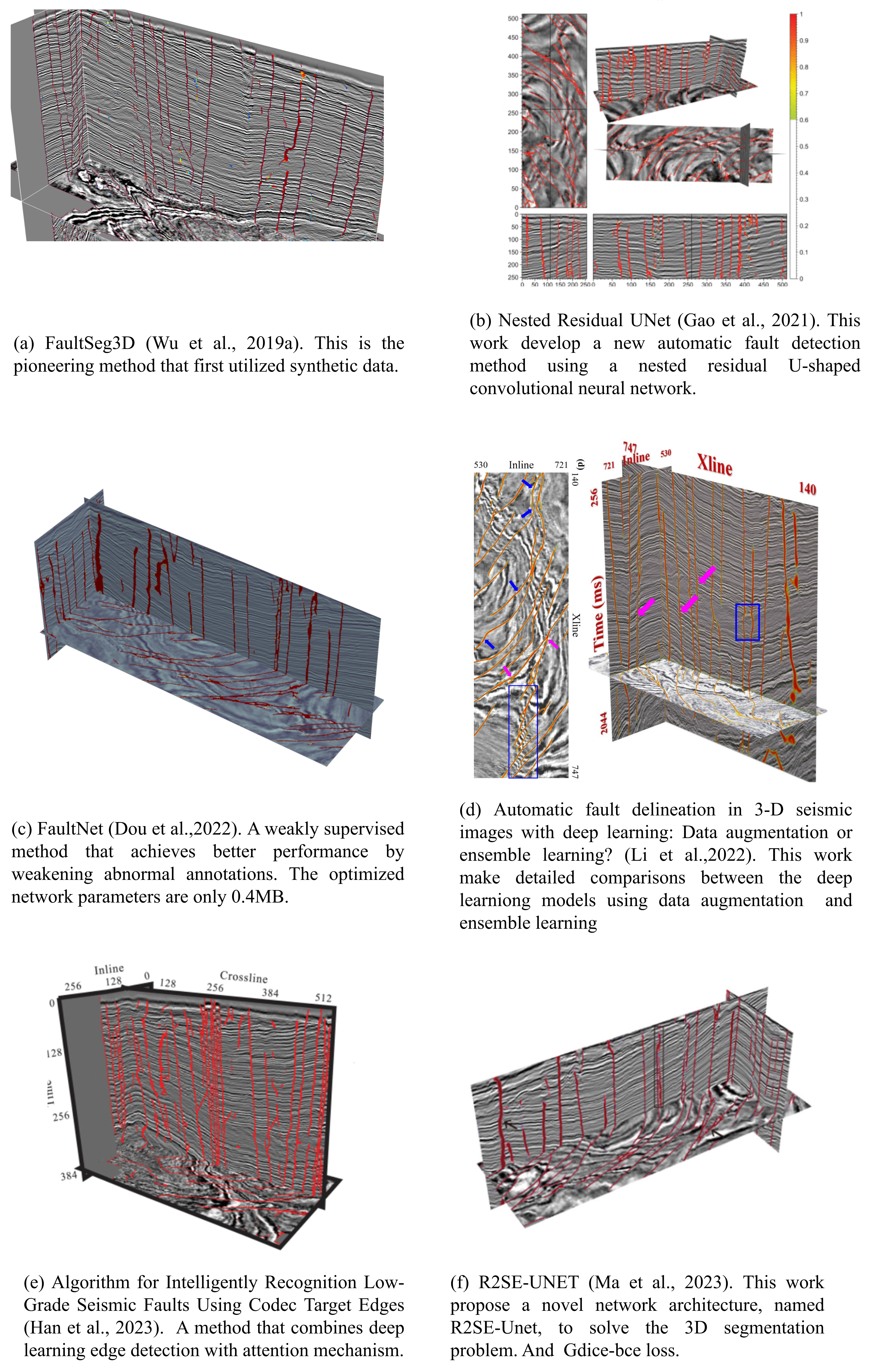}
	\centering\caption{Appendix figure: The detection results of New Zealand's Kerry from all papers published in the two authoritative journals, Geophysics and TGRS, from 2019 to 2023. We summarized the main innovations of each method in one sentence.}
	\label{apd2}
\end{figure*}

This appendix summarizes the detection results of two classic surveys, F3 and Kerry, from methods published in two authoritative journals after 2019(Geophysics an TGRS, Figure \ref{apd1-1}, \ref{apd1-2}, \ref{apd2}). These results are arranged in chronological order and reflect the development of fault detection over these four years to a certain extent. These methods are respectively \cite{wu2019faultseg3d,feng2021uncertainty,dou2021attention,dou2022md,gao2022automatic,alohali2022automated,han2023algorithm,kaur2023deep,ma20233d,10196009,gao2021fault,li2022automatic}.

\bibliographystyle{IEEEtran}
\bibliography{example}

\end{document}